\documentclass[%
  ]{article}
\usepackage[a4paper, total={6in, 8in}]{geometry}
\usepackage{array}
\usepackage{setspace}
\usepackage{graphicx} 
\usepackage{authblk}
\usepackage{graphicx}
\usepackage{dcolumn}
\usepackage[usenames,dvipsnames]{color} 
\usepackage{bm}
\usepackage{amsmath} 
\usepackage{amssymb}
\usepackage{parskip}
\usepackage{blindtext}
\usepackage{moresize}
\newcommand{\avg}[1]{\left\langle #1 \right\rangle}
\newcommand{\abs}[1]{\left\lvert#1 \right\rvert}

\newcommand{\vrr}{\mathbf{r}}
\newcommand{\vk}{\mathbf{k}}
\usepackage{hyperref}
\DeclareUnicodeCharacter{2061}{}
\begin{document}
\title{Mirror Symmetry in three-dimensional Multiple-Scattering Media\\
}

\author[1]{Sudhir K. Saini}
\author[1,2]{Evangelos Marakis}
\author[1]{Kayleigh Start}
\author[3]{Gerwin Osnabrugge}
\author[3]{Ivo M. Vellekoop}
\author[1]{Pepijn W.H. Pinkse}
\affil[1]{MESA+ Institute for Nanotechnology, University of Twente, The Netherlands}
\affil[2]{Institute of Electronic Structure and Laser, Foundation for Research and Technology-Hellas, Crete, Greece}

\affil[3]{TechMed Centre, University of Twente, The Netherlands}
\date{}
\maketitle

\begin{abstract}We investigate the effect of a mirror-symmetry plane in multiple-scattering media under plane-wave illumination along the symmetry plane. Designed and fabricated samples’ optical transport properties are compared quantitatively with three-dimensional modeling. Strong polarization-dependent deviations of the bulk speckle-averaged intensity distribution at the symmetry plane are observed, showing either up to a factor two enhancement or complete suppression of the ensemble-averaged intensities. We derive analytical expressions for the ensemble-averaged intensity profiles near the symmetry plane. Apart from their interest to fundamental light propagation studies, applications of mirror-symmetric scattering media are envisioned in anti-counterfeiting.
\end{abstract}

Symmetry is a fundamental concept in describing the physical laws of the universe [1]. The intricate geometrical symmetries found in many living organisms have inspired researchers to explore the use of symmetries in nanophotonic media for controlling light propagation [2]. At first glance, it seems that symmetry and randomness are mutually exclusive concepts. Indeed, random materials created by nature lack global symmetries like point or mirror symmetry. However, man-made structures may be designed to exhibit both local randomness and global symmetry, providing an opportunity to study and alter light transport in new ways. Naturally occurring or man-made random photonic structures have been investigated from the light localization perspective, with possible applications in random lasers and photovoltaic devices, where they can enhance light trapping [3-8]. These studies have largely relied on the statistical properties of light transport, as the structures form through some sort of self-assembly process. Lately, it has been shown that it is possible to deliberately engineer the spatial correlation and the degree of disorder within the sample to produce intriguing interference effects [9,10]. Some studies emphasized controlling light transport using bio-inspired spatially correlated photonic structures, which allows controlling the propagation of light [11-13]. 
Recent theoretical work has shown that constructive interference between mirror-symmetric classical electron paths significantly increases conductance through an opaque barrier placed within a symmetric quantum dot [14,15]. In the microwave regime, it has been demonstrated numerically and experimentally that mirror symmetry can cause a broadband improvement of transmission across an opaque barrier when illuminated perpendicular to the symmetry plane [16,17]. However, we are not aware of any exploration of mirror-symmetric three-dimensional (3D) scattering media illuminated in a symmetry or antisymmetric way in the optical wavelength range. Remarkable progress has been made in the field of nanofabrication techniques, such as electron-beam lithography and direct laser writing, which have enabled the production of pre-designed scattering media with submicron accuracy [18,19]. Consequently, deterministic scattering media with features on the length scale of the optical wavelength can now be fabricated.
In this work, we investigate the striking behavior of polarized light in a mirror-symmetric disordered multiple-scattering medium. We fabricate mirror-symmetric samples by direct laser writing, show optical transport data obtained with them, and compare them with theoretical and numerical modeling. To numerically investigate the wave propagation in our mirror-symmetric dielectric multiple-scattering media, we employ modified-Born-series [20] and finite-difference time-domain methods. Our investigation, combining numerical simulations, theoretical analysis based on field-field correlation functions, and experimental measurements, revealed consistent polarization-dependent behavior of the ensemble-averaged intensity distribution of a bulk system near its symmetry plane. Moreover, we find that the ensemble-averaged intensity as a function of the distance to the symmetry plane can also be predicted by the analogy to the situation studied by Drexhage, namely a dipole emitter close to a mirror.
To illustrate our system, we present a mirror-symmetric two-dimensional (2D) configuration in Fig. 1. The model consists of point scatterers distributed randomly in the y-z plane. The random scatterers are mirrored in the x-z plane, at y = 0. They are illuminated by an incident plane wave with propagation direction z, parallel to the symmetry plane, preserving the symmetry of the system. For each path from the light source to a point on the symmetry line there is a mirror-symmetric copy with an equal path length. At the symmetry plane, this ray and its mirror copy will interfere. Since the path lengths are equal, their phases will be identical, causing constructive interference and an intensity at the symmetry plane that is higher than the average value on either side. Naturally, to observe this constructive interference experimentally, optical pathways from illumination to and through the sample must have the same length to better than a quarter of the wavelength to achieve the expected interference.
  
\begin{figure}[!htb]
\includegraphics[width=1\linewidth]
{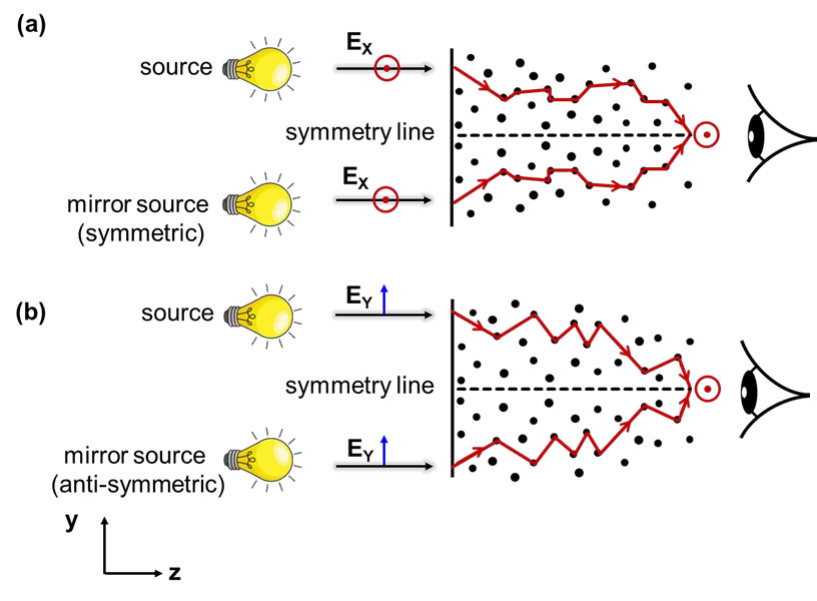}
\centering
\label{fig:enter-label}
\caption{ Illustration of a mirror-symmetric medium with point scatterers illuminated by a plane wave. Every pathway from a point in the illumination to a point on the symmetry line (dashed line) has a mirror-symmetric counterpart, whereas for points away from the symmetry plane there is no such counterpart. The scatterers are illuminated parallel to the symmetry plane by light linearly polarized (a) parallel ($E_X$, red circle) and (b) perpendicular ($E_Y$, blue line) to the symmetry plane. However, at the output, only the parallel x-polarization is observed. }

\label{fig:enter-label}
\end{figure}

Since light has a vector nature, one also needs to consider the polarization of the incident light. Figs. 1(a) and 1(b) show our mirror symmetric configuration visualizing the polarization degree of freedom: The sample is illuminated with a polarized plane wave propagating in the positive z-direction. The polarization of the incident field is chosen parallel ($E_X$) and perpendicular ($E_Y$) with respect to the symmetry plane in Figs. 1(a) and 1(b), respectively. In the case where both the input and measured polarizations are aligned along the x-axis (hereafter referred to as XX polarization), the system's overall symmetry is preserved." and we expect a factor 2 higher intensity on average due to constructive interference on the symmetry line. Contrarily, changing the input polarization to perpendicular along the y-axis creates an antisymmetric input state, leading to destructive interference in the x-polarization at the symmetry plane and to a distinct dark line on the symmetry plane, as predicted by our theory. Qualitatively, this cancellation is also expected for the scenario where the incident polarization is along the x-axis, but the perpendicular polarization along Y (hereafter referred to as XY polarization) is measured. Likewise, constructive interference at the symmetry line is expected in the YY case.
Further we compute the ensemble-averaged intensity distribution at the sample's rear surface. We have constructed an analytical model for our mirror-symmetric system (using field-field correlation functions, see supporting information section I). We examined scalar waves first, followed by vector waves. The scalar-wave theory always predicts intensity enhancement at the symmetry plane. Only the full vector theory predicts the dark interference line at the symmetry plane for some polarization configurations. Remarkably, the vector theory predicts much richer behavior absent in the scalar theory. It can be noted that the scalar predictions cannot be understood as an average over the different polarization cases. We divide the phenomenon into four categories based on the polarization of the incident light and the observation polarization at the sample's exit surface in order to thoroughly analyze the phenomenon. The observed intensities are for the four cases

\begin{equation}
    I_{XX}(d)=1+\frac{3\sin⁡(d)}{2d}+\frac{3\cos⁡(d)}{2d^2}-\frac{3\sin⁡(d)}{2d^3}         
\end{equation}
\begin{equation}
    I_{XY}(d)=1+\frac{3\cos⁡(d)}{d^2}-\frac{3\sin⁡(d)}{d^3}   
\end{equation}
\begin{equation}
    I_{YY}(d)=1-\frac{3\cos⁡(d)}{d^2}+\frac{3\sin⁡(d)}{d^3} \end{equation}
\begin{equation}
    I_{YX}(d)=1-\frac{3\sin⁡(d)}{2d}-\frac{3\cos⁡(d)}{2d^2}+\frac{3\sin⁡(d)}{2d^3}         
\end{equation}

where $d=4{\pi}n_{eff} y/\lambda$, $n_{eff}$ is the effective refractive index of the medium, y is the distance to the symmetry plane, and $\lambda$ is the wavelength of the incident light.

To validate the above predictions in 3D, we do a numerical simulation of light propagating along the positive z-axis (see supporting information section II). Figure 2 shows the simulation results of a 3D mirror-symmetric sample consisting of randomly filled voxels (0.125 µm × 0.125 µm × 0.125 µm) with a fill fraction of 50, acting as scatters. The sample has a symmetry plane parallel to the x-z-plane, positioned at y = 0. A parallelly polarized plane wave is injected along the positive z-axis. The scattered intensity is recorded at the sample's x-y plane. Figures 2(a) and 2(b) show 2D maps of the symmetric speckle patterns averaged over the z-direction. As expected, a bright line appeared at the symmetry plane for the $I_{XX}$ component of the transmitted light. In contrast, a minimum intensity line is observed at the symmetry plane for the $I_{XY}$ component of the transmitted light. These results corroborate the polarization-dependent interference effects as a consequence of the symmetry of the sample. We have also observed a polarization-dependent intensity at the symmetry plane when the input light is polarized perpendicular to the symmetry plane: $I_{YX}$ shows a dark line and $I_{YY}$ a bright line at the symmetry plane. The differences between the XX vs the YY cases and the XY vs YX are small, but the numerical simulations again follow the analytic formulas.
\begin{figure}[!htb]
\centering
\makeatletter 
\renewcommand{\thefigure}{\@arabic\c@figure}
\makeatother
\includegraphics[width=0.75\linewidth]{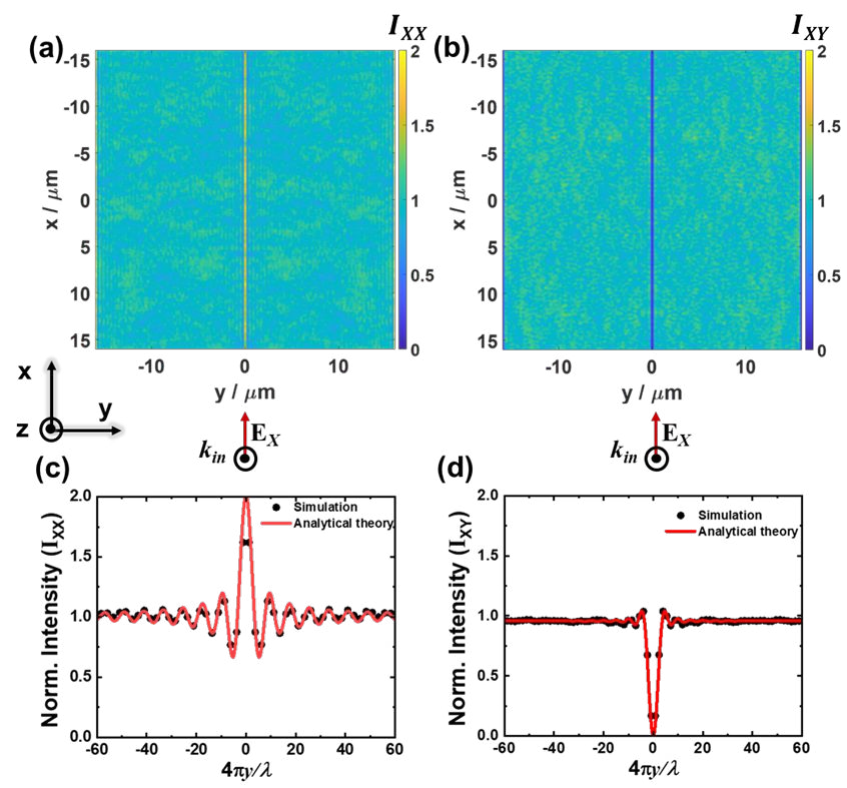}
\caption{ 3-Dimensional numerical simulation results of mirror-symmetric scattering media, with the input wave polarized perpendicular to the symmetry plane. The 2D maps of the 3D symmetric speckle patterns averaged over the z-direction show (a) a maximum intensity line at the symmetry plane for $I_{XX}$ and (b) a minimum-intensity line at the symmetry plane for the $I_{XY}$ component of the transmitted light. The intensity profile obtained from the 2D maps by averaging over x-dimension (symbols) are plotted in (c) for the $I_{XX}$ component and in (d) for the $I_{XY}$ component of the output light. The curves are fitted (line) with our analytical model with $n_{eff}$ as a fitting parameter. Note that the intensity of the scattered light for the $I_{XX}$ component is 1.88 times higher than that of the $I_{XY}$ component of the transmitted light.}
\label{fig:enter-label}
\end{figure}
Figures 2(c) and 2(d) show the intensity line profile (symbols) for the $I_{XX}$ and $I_{XY}$ components of the transmitted light averaged over the x-direction, respectively. The $I_{XX}$ component peaks, whereas the $I_{XY}$ component dips to a low value at the symmetry plane y = 0. Mainly around the peak in (2c) but also around the dip in (2d) fringes can be seen. Figure 2(c) and 2(d) show the intensity line profile of simulated data fitted using equations 1 and 2, respectively. Using $n_{eff}$ as free parameter, a good fit is obtained for $n_{eff}$ = 1.25. This value matches the $n_{eff}$ value calculated using scalar effective medium theory [13]. From this, we conclude that in our analytic theory (derived in the Supplement) is a good description of our system. While our analytic approximation predicts the peak/dip (2/0) values, it's notable that we don't entirely reach those values in our simulation results. This discrepancy occurs because the theoretical results assume infinite slabs, whereas our samples are small and finite. As a result, a slight deviation from the fit is expected, underscoring the finite nature of our samples compared to the theoretical assumptions of infinite slabs. Moreover, we have also noticed the resemblance of our mirror-symmetric system with the configuration of a single dipole emitter above a mirror as considered by Drexhage [21,22]. Assuming the density of scatterers in the vicinity of the symmetry plane is low, we can approximate the average field around the symmetry plane by taking the expressions from Drexhage for a single emitter above a mirror without any additional phase at its surface. The results reproduce the rigorous analytical results based on the field-field correlations. Since both theories are an approximation to some point, we have also performed numerical simulations to gain more insight into the intensity line profile variations of the mirror-symmetric scattering media as a function of scattering mean free path ($l_s$) (see Supporting information section III). From these simulations, we learned that the sample dimensions should be large compared to the scattering length for the approximation to be good. 
\par
Additionally, we conducted a thorough analysis of the sensitivity of the optical properties of mirror-symmetric media to manufacturing imperfections. This analysis highlights how such imperfections affect the symmetry of the sample. We modified the equation to include a parameter$\gamma$, which ranges from 0 for a completely disordered medium to 1 for perfect symmetry (see supporting information section IV). This investigation focuses on how small perturbations in the arrangement of the scatterers affect the overall optical behavior of the system. So here, we consider two scenarios of perturbation within a mirror-symmetric medium consisting of spheres with a 200 \textit{nm} radius. In the first scenario, we perturbed a portion of our scatterers by distances chosen from a Gaussian distribution with a mean of zero and a standard deviation of 200 \textit{nm} from their original positions. We then compared the resulting speckle patterns for both polarizations with those from a non-perturbed ensemble. In the second scenario, we perturbed all scatterers in the medium, applying perturbations drawn from Gaussian distributions with standard deviations incrementally increasing from 40 \textit{nm} to 200 \textit{nm}, to assess the optical response for both polarizations. Statistical averages are taken from five realizations in each scenario. The output intensities and refractive index (RI) distributions are compared using Pearson correlations, as shown in Figure 3, and the $\gamma$ field correlation (see supporting information Figure S2). The results indicate that mirror-symmetric media are highly sensitive to fabrication errors. However, some symmetric interference effects are still observable, even with perturbations, given a fabrication positional accuracy of up to 20 \textit{nm}. Specifically, we found that the Pearson correlation falls below 50\% once the average displacement exceeds approximately one-tenth of the wavelength (60 nm for a 600 \textit{nm} wavelength). Thus, while fabrication errors can significantly impact the optical properties of mirror-symmetric media, our fabrication technique, with a positional error margin of 20 \textit{nm}, remains feasible for observing the expected peaks and dips as predicted by theory. This suggests that our method can reliably produce the designed optical effects despite minor fabrication imperfections.
\begin{figure}[!htb]
\centering
\makeatletter 
\renewcommand{\thefigure}{\@arabic\c@figure}
\makeatother
\includegraphics[width=0.75\linewidth]{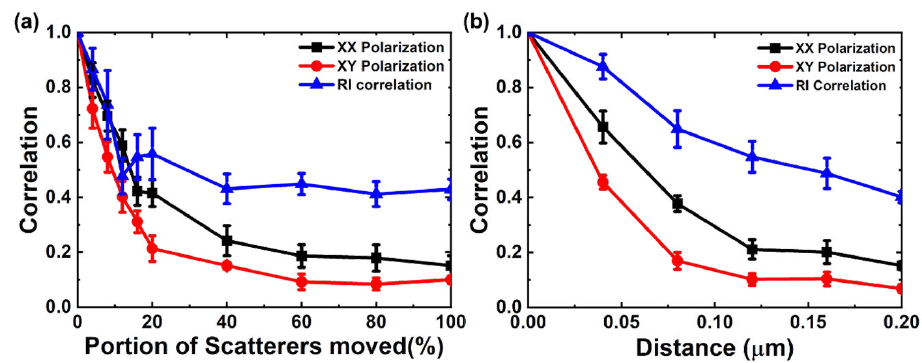}
\caption{ Correlation of optical properties with increasing perturbation. This figure illustrates how the correlation between the intensity patterns and refractive index distributions of perturbed systems and an unperturbed reference evolves with increasing perturbation magnitude. (a) The intensity correlation of XX and YY polarizations, along with the refractive index distribution correlation, is plotted against the portion of scatterers perturbed by distances from a Gaussian distribution with a standard deviation of 200 $nm$ in the mirror-symmetric medium. Error bars denote the standard deviation, while the solid lines represent the ensemble average values. (b) Intensity and refractive index correlations for scenarios where all scatterers are perturbed by varying standard deviations, revealing that the Pearson correlation falls below 50\%. }
\label{fig:enter-label}
\end{figure}
\par
In order to experimentally verify the predictions of our models, we have fabricated deterministic mirror-symmetric scattering media using a direct laser writing method. We have designed mirror-symmetric scattering media using an algorithm based on Jaynes' solution to Bertrand's paradox, providing the random placement of rods in a cubic volume with an on-average-homogeneous filling of the volume [23]. A set of random but on-average-homogeneously distributed points is produced on a sphere using a random number generator, and line segments (“chords”) are generated by connecting each pair of points. Finally, the structure with the desired dimension is cut out from the sphere. Further, these designed structures are mirrored by flipping them in the y-direction and making them mirror-symmetric random media. Figure 4(a) shows the design model of a scattering medium with 200 randomly oriented rods filling a cube of 15 µm × 15 µm × 5 µm. We can control the number of rods in a given volume, which provides control over the fill fraction of polymer, leading to the desired scattering strength of the disordered medium. Using the direct writing laser method, we have used the same design model to fabricate it on a micro-glass slide (see Supporting information section V). Figure 4(b) depicts a SEM image of a produced structure based on the design model. In the top view, the symmetry is clearly visible. The estimated rod thickness is 524 ± 60 \textit{nm}. It is observed that the design and the actual structure are very similar, and the features are in the same relative positions. It can be noticed, however, that the synthesized structure has a few errors, such as missing rods and shrinkage artifacts, which emerge during the developing process. The overall number of missing rods is minimal.We have also made several copies of a similar design using varying writing speeds (See Supporting Information section V). This further supports that our method is more reliable for producing identical copies of mirror symmetric scattering media.    

\begin{figure}[!htb]
\centering
\makeatletter 
\renewcommand{\thefigure}{\@arabic\c@figure}
\makeatother   
\includegraphics[width=0.75\linewidth]
{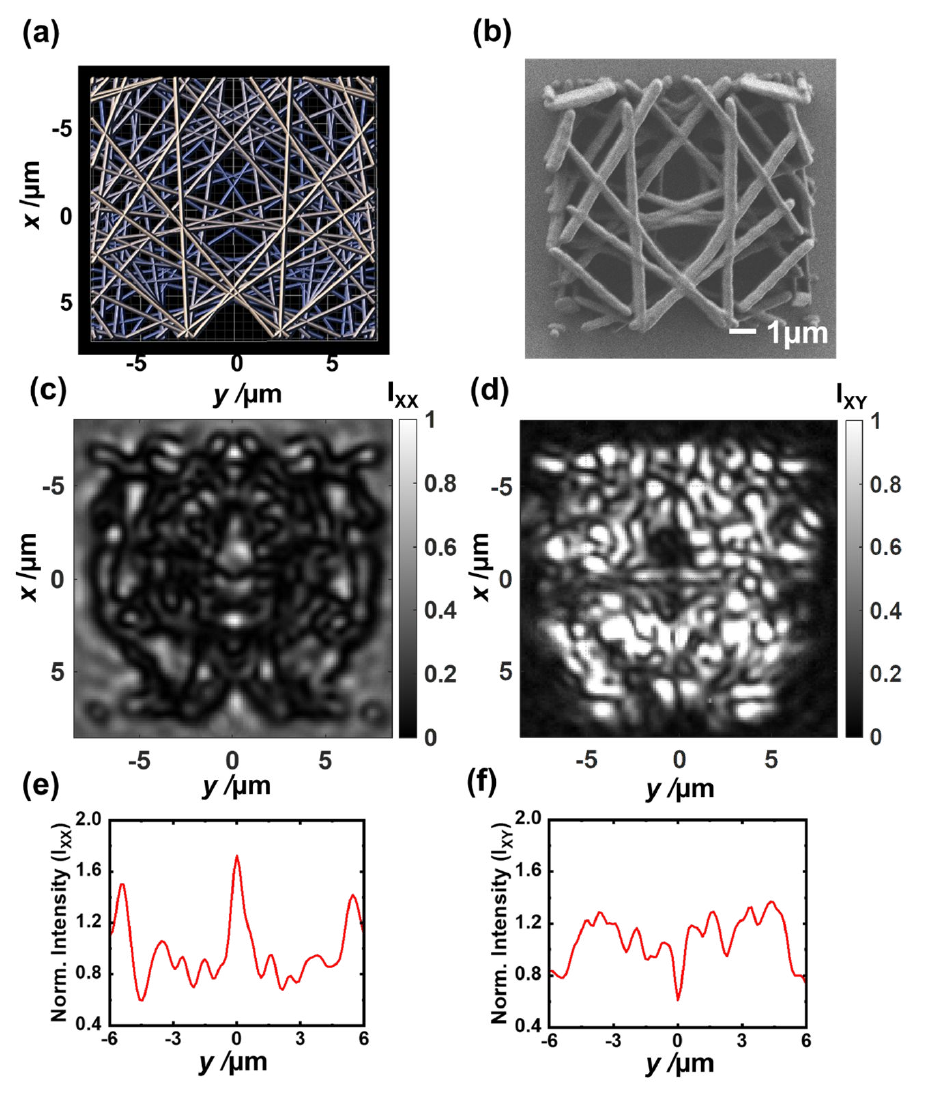}
\caption{ (a) Design model of a mirror-symmetric scattering medium with randomly oriented rods filling a cube. (b) SEM image showing fabrication result of the design model in (a), fabricated using direct laser writing, scale bar 1 µm. Light transmission measurements through mirror-symmetric scattering sample with input light polarized parallel (EX) to the symmetry plane are shown in (c-d). The measured speckle patterns in (c) show the $I_{XX}$ component and in (d) the $I_{XY}$ component of the output light. Intensity line profile summed over the x values are shown in (e) and (f) for the two respective components, where the units are the summed grey values of the camera pixels normalized to the average intensity of the speckle originating from the bulk of the structure.}
\label{fig:enter-label}
\end{figure}
A home-built setup is used to measure the light transmission through the mirror-symmetric samples (see Supporting Information section VI). Figures 4(c) and 4(d) show the measured intensity patterns from the fabricated cubes for the $E_X$ and $E_Y$ components of the transmitted light, respectively. In the case of the $E_X$-component of the transmitted light, as expected, a high-intensity line from each cube is observed at the symmetry plane when these structures are illuminated with perpendicular polarized light to the symmetry plane. In contrast, a dark line is observed at the symmetry plane for the $E_Y$-component of the transmitted light. Figures 4(e) and 4(f) show the intensity line profiles of the output light's $E_X$-component and $E_Y$-component, respectively. These are obtained by summing Figs. 4(c) and (d) over the x-direction. At the symmetry plane y=0, a peak for the $E_X$-component and a dip for the $E_Y$-component can be seen clearly. Clear periodic fringes around the peak, as in the simulated Figs. 2, are absent in our measured images. We hypothesize that the lacking fringes is due to the finite NA of the objective and the finite sample thickness. As per our numerical calculations, we estimate that the ls within the presented structures spans a range of 2 to 4 µm. This observation is in agreement with the scattering regime depicted in Fig. S1 (see Supporting information section III), where only the central interference peak prevails, exhibiting a peak-to-background ratio consistent with the reported values. It is also important to consider the potential impact of fabrication imperfections on our analysis. Such errors may disrupt the mirror symmetry of the structural features, thereby influencing the corresponding image and potentially altering the observed scattering phenomena.
It is important to highlight that the findings of this study can be extrapolated to any type of wave propagation, be it acoustic, matter, or electromagnetic waves, given the appropriate length scales. Moreover, at the symmetry plane the optical density of states will also be altered by the symmetry, presenting opportunities for the investigation of quantum electrodynamics phenomena, such as controlled emission through single emitters. In the field of metrology, mirror-symmetric systems may be established as proficient mechanisms for aligning masks in multi-layer lithography procedures and evaluating the manufacturing quality of the associated instrument.
\par
In conclusion, we have demonstrated the effect of mirror symmetry on light propagation in dielectric multiple-scattering media. Analytic modelling predicts the ensemble-averaged intensity distribution around the symmetry line. The modeling, the numerical simulations and the optical characterization results establish large polarization-dependent deviations at the symmetry plane of the bulk ensemble-averaged intensity distribution. Although the mirror-symmetric media are highly sensitive to scatterer perturbations, our fabrication technique with a 20 \textit{nm} accuracy can still reliably produce the designed optical effects despite minor imperfections. The current investigation has unveiled novel avenues for exploring fundamental principles of light propagation, devising anti-counterfeiting measures, detecting manufacturing defects through wave-based techniques, and implementing non-destructive testing methodologies..\\
{\textbf{Acknowledgements:}}This publication is part of the project ‘Free form scattering optics’ (FFSO) with project number P15-36 of the TTW Perspectief research programme which is financed by the Dutch Research Council (NWO). I.M.V. acknowledges support from ERC PoC grant 101069402 (WAVESIM). We also thank Daan P. Stellinga, Oluwafemi S. Ojambati, Mohammad R. Aghdaee, and Ad Lagendijk for fruitful discussions.

\newpage
\renewcommand\thesection{\Roman{section}}
\renewcommand\thesubsection{\thesection.\Roman{subsection}}

               {\textbf{Supporting Information}}\\
This Supplementary Information discusses details of the numerical modelling as well as a rigorous derivation of the average intensity distribution of light scattering in mirror-symmetric media. Further, we discuss the similarities of our model with the Drexhage model for a point emitter close to a mirror. At the end we describe sample fabrication and the optical measurement setup.

\section{Rigorous derivation of the intensity distributions}
We consider transmission of light through a slab of scattering material extending infinitely in the $x$ and $y$ directions. The structure has a mirror symmetry with symmetry plane $y=0$. The slab is illuminated by a plane wave travelling in the positive $z$ direction, i.e. normal to the surface, and we are interested to compute the ensemble-averaged intensity distribution at the back surface of the sample.

We start analyzing the problem for scalar waves and then modify the approach for vector waves. Surprisingly, the vector waves' behavior is quantitatively and qualitatively very different from that of the scalar waves.

\subsection{Scalar waves}
Let $E(x,y)$ be the field at the back surface of the structure. We separate the field into two contributions: let $E_L(x,y)$ be the transmitted field resulting from illuminating only the left half of the sample ($y<0$), and let $E_R(x,y)$ be the transmitted field resulting from illuminating only the right half of the sample. 

Because of the mirror symmetry in the $y=0$ plane, we have 
\begin{equation}
    E_R(x,y) = E_L(x,-y)\label{eq:LR-start}
\end{equation}
Illuminating both halves at the same time gives the
intensity distribution
\begin{align}
    I(x,y) &= \abs{E_L(x,y) + E_R(x,y)}^2\\
    &=\abs{E_L(x,y)}^2 + \abs{E_L(x,-y)}^2\nonumber\\
    &\quad+ 2\Re\left[E_L(x,y)^* E_L(x,-y)\right]
\end{align}
We proceed to compute the ensemble average of this expression. Assuming that $x,y\ll L$, 
{ where $L$ is the thickness of the slab,} 
we can approximate $\avg{\abs{E_L(x,y)}^2}\approx \avg{\abs{E_L(0,0)}^2} \equiv I_0/2$ and find
\begin{align}
    \avg{I(x,y)} &= I_0 \left[1 + \Re\;C(2\Delta y)\right],\label{eq:scalar-intensity}
\end{align}
with $C$ the (ensemble averaged) field correlation function. This function is invariant along $x$. Since $x,y\ll L$, we can approximate it as 
\begin{equation}
C(\Delta y) \equiv \frac{\avg{E_L^*(0, -\Delta y/2) E_L(0, 0+\Delta y/2)}}{\sqrt{\avg{|E_L(0, 0-\Delta y/2)|^2}\avg{|E_L(0, 0+\Delta y/2}|^2}}
\end{equation}
We further assume that the symmetry plane does not affect the correlations in $E_L$ or $E_R$ separately, so we can just use the correlation function $C$ as computed for a scattering medium without a symmetry plane. Since $C(0)\equiv 1$, Eq.~\eqref{eq:scalar-intensity} predicts that the average intensity at the mirror plane will be twice that of the average intensity elsewhere.

To compute $C$, we consider a plane-wave decomposition of the field $E$. Denote by $\xi(\vk)$ the random complex amplitude of a plane wave leaving the sample in direction $\vk$, so that the transmitted field is given by:
\begin{equation}
    E(\vrr) = \iint d\vk \xi(\vk) e^{i \vk\cdot\vrr}
\end{equation}
with $\vrr = (x,y,z)$, and the integral is taken over all wave vectors leaving the sample ($2\pi$ solid angle). For a fully developed speckle pattern in the far field, we have the properties $\avg{\xi(\vk)}=0$ and $\avg{\xi^*(\vk_1)\xi(\vk_2)}=\delta(\vk_2-\vk_1)I(\vk_1)$, with $I(\vk_1)$ the intensity distribution (i.e., the radiant intensity) of light propagating in direction $\vk$. The correlation function at the $z=0$ plane can be computed as

\begin{multline}
C(\Delta y)=\frac{1}{I_0}\avg{E^*(0,-\Delta y/2) E(0,\Delta y/2)} =\\
\frac{1}{I_0}\iint d\vk_1 \iint d\vk_2 \avg{\xi^*(\vk_1)\xi(\vk_2)} e^{i\left[k_{1y} \Delta y/2 + k_{2y}\Delta y/2)\right]}\\
= \frac{1}{I_0}\iint d\vk I(\vk) e^{i k_y \Delta y}
\end{multline}
which is a form of the van Cittert-Zernike theorem. We now choose spherical coordinates, with $\theta=0$ corresponding to the $y$ axis, and consider a Lambertian surface ($I(k) = I_0/(2\pi)$ = constant).

\begin{align}
C(\Delta y) &= \frac{1}{2\pi} \int_0^{\pi} d\phi \int_0^{\pi} \sin(\theta) d\theta e^{i k \Delta y\cos(\theta) }\\
&=\frac{\sin(k \Delta y)}{k \Delta y}
\end{align}
where $k=\lVert \vk \rVert = \frac{2\pi n}{\lambda}$ with $n$ the effective refractive index of the medium and $\lambda$ the wavelength of the light. 

\subsection{Vector waves}
For vector waves, the situation is somewhat different since the field has three vector components. The situation is still symmetric if we illuminate with $x$, or $z$-polarized light and observe transmitted $x$ or $z$-polarized light. However, if either the incident or the observed light is $y$ polarized, the situation is different since mirroring the field in the $y=0$ plane introduces a sign change. Therefore, for those situations, we have $E_R(x,y) = - E_L(x,y)$. If both incident and observed light is $y$-polarized, the two minus signs cancel and again we have $E_R(x,y) = E_L(x,y)$ 

A second difference is that the radiant intensity coming from the surface will depend both on the angle and on the polarization. Therefore, the associated correlation functions will be different for different polarizations. The theoretical intensity profiles for all combinations of incident and transmitted light are summarized in Table \ref{tab:intensity-functions}, where $C_x$, $C_y$, and $C_z$ are the correlation functions of the field at the back surface of the sample for a \emph{non}-symmetric structure. Note that there cannot be a $z$-polarized incident beam since we assume propagation by a plane wave propagating in the $z$-direction. With a high numerical aperture microscope objective, however, it will be possible to collect some of the light that is $z$-polarized (i.e. axially polarized) at the sample surface. This light will split evenly over the $x$ and $y$ polarization detection channels (see columns $x$-camera and $y$-camera in Table \ref{tab:intensity-functions} for the case of a numerical aperture of 1).

\begin{table*}
\begin{tabular}{ l |c|c|c|c|c } 
 & out=$x$ & out=$y$ & out=$z$ & out=$x$-camera & out=$y$-camera\\
 \hline
 in = $x$ & $1 + C_x(2\Delta y)$ & $1 - C_y(2\Delta y)$ & $1 + C_z(2\Delta y)$ & $1 + \frac23 C_x(2\Delta y) + \frac13 C_z(2\Delta y)$ & $1 - \frac23 C_y(2\Delta y) + \frac13 C_z(2\Delta y)$ \\ 
 in = $y$ & $1 - C_x(2\Delta y)$ & $1 + C_y(2\Delta y)$ & $1 - C_z(2\Delta y)$ & $1 - \frac23 C_x(2\Delta y) - \frac13 C_z(2\Delta y)$ & $1 + \frac23 C_y(2\Delta y) - \frac13 C_z(2\Delta y)$ \\ 
 in = $z$ & n/a & n/a & n/a & n/a & n/a \\ 
\end{tabular}
\caption{Theoretical intensity profiles of the average transmission through a scattering structure with mirror symmetry in $y=0$, for all combinations of incident and transmitted light. All profiles are relative to the transmission through a non-symmetric structure. Correlation functions $C_y$, and $C_x = C_z$ are given by equations \eqref{eq:C_y} and \eqref{eq:C_x}, respectively. }
\label{tab:intensity-functions}
\end{table*}

We now proceed to compute $C_x$, $C_y$ and $C_z$, i.e. the correlation functions of the transmitted light in the \emph{non-symmetric case}. To compute the correlation functions, we assume that the field is generated by a set of randomly oriented dipole emitters, emitting at random amplitude and phase. Each dipole can be decomposed into (statistically independent) $x$, $y$, and $z$ components.

Like in the scalar case, denote by $\xi_i(\vk)$ the complex amplitude of a plane wave originating from a dipole oriented in direction $i\in x,y,z$. Again, we can define the propagating field as
\begin{equation}
    E(\vrr) = \iint d\vk \xi_i(\vk) e^{i \vk\cdot\vrr},
\end{equation}
giving the correlation function $z=0$ plane
\begin{multline}
C(\Delta y)=\frac{1}{I_0}\avg{E^*(0, -\Delta y/2) E(0, \Delta y /2)} =\\
\frac{1}{I_0}\iint d\vk_1 \iint d\vk_2 \avg{\xi_i^*(\vk_1)\xi_i(\vk_2)} e^{i\left[k_{2y} \Delta y/2 + k_{1y}\Delta y/2\right]}\\
= \frac{1}{I_0}\iint d\vk I_i(\vk) e^{i k_y \Delta y}
\end{multline}

For dipole radiation, the radiant intensity $I_i(\vk)$ depends both on the polarization and the propagation angle. We start by computing $C_z$ and choose spherical coordinates, with $\theta=0$ corresponding to the $z$-axis and $\theta=\pi/2, \phi=0$ corresponding to the $y$-axis. In these coordinates, the dipole radiation pattern for a $z$-polarized source is given by $I_z(\theta) = \frac{3 I_0} {4\pi} \sin^2\theta$, giving
\begin{align}
C_z(\Delta y) &= \frac{3}{4\pi}\int_0^{2\pi} d\phi \int_0^{\pi/2} \sin(\theta) d\theta \sin^2(\theta) e^{i k \Delta y\sin(\theta)\cos(\phi) }\\
&=\frac{3}{2}\left(\frac{\sin(k \Delta y)}{k \Delta y}+\frac{\cos(k \Delta y)}{k^2 \Delta y^2}-\frac{\sin(k \Delta y)}{k^3 \Delta y^3}\right)\label{eq:C_z}
\end{align}
where $I_0 = 3/(4\pi)$ is a normalization factor such that $C_z(0)=1$. 

For the $x$ polarization, we rotate the coordinate system to align with the $x$-axis (placing the $y$-axis at $\theta=\pi/2, \phi=0$), and adjust the integration limits to still correspond to the $2\pi$ steradian of outgoing angles.
\begin{align}
C_x(\Delta y) &= I_0 \int_0^{\pi} d\phi \int_0^{\pi} \sin(\theta) d\theta \sin^2(\theta) e^{i k \Delta y \sin(\theta)\cos{\phi}}\\
&=\frac{3}{2}\left(\frac{\sin(k \Delta y)}{k \Delta y}+\frac{\cos(k \Delta y)}{k^2 \Delta y^2}-\frac{\sin(k \Delta y)}{k^3 \Delta y^3}\right)\label{eq:C_x}
\end{align}
which is exactly the same as for $z$-polarization.

Finally, for $y$-polarization, we align the coordinate system so that $\theta=0$ corresponds to the $y$-axis. 
\begin{align}
C_y(\Delta y) &= I_0 \int_0^{\pi} d\phi \int_0^{\pi} \sin(\theta) d\theta \sin^2(\theta) e^{i k \Delta y \cos(\theta)}\\
&=3\left(-\frac{\cos(k \Delta y)}{k^2 \Delta y^2}+\frac{\sin(k \Delta y)}{k^3 \Delta y^3}\right)\label{eq:C_y}
\end{align}

These correlation functions, when used with the equations in Table \ref{tab:intensity-functions}, exactly reproduce the functions for the emission rate of a fluorescent molecule in the vicinity of a mirror, as derived by Drexhage et al. \cite{Drexhage1968, Drexhage1970}. This is not surprising, since those functions also describe a phenomenon that is directly the result of interference of a dipole with its mirror image.

To take into account the effect of far-field detection when measuring the light escaping from a sample, we have added two columns to Table \ref{tab:intensity-functions}
where the effect of an objective with an NA=1 is assumed. Such an objective collects 50\% of the escaping light, independent of the  polarisation (x,y,z). Because of the symmetry of the system, x (y) polarization at the sample surface will lead to x (y) polarization in the camera. The z polarization will be distributed 50\% - 50\% over both channels. Assuming every polarization has the same Lambertian intensity, you get, after normalisation, the factors 1/3 and 2/3 from the table.

\subsection{Correction for ballistic transmission}
In the above, we used a random distribution $\xi$ with the properties $\avg{\xi(\vk)}=0$ and $\avg{\xi^*(\vk_1)\xi(\vk_2)}=\delta(\vk_2-\vk_1)I(\vk_1)$. The earlier assumptions that $\avg{\abs{E_L(x,y)}^2}\approx \avg{\abs{E_L(0,0)}^2}$ and $C(x,y,\Delta y)\approx C(0,0,\Delta y)$ follow directly from the use of this distribution.

This distribution corresponds to a fully developed speckle pattern. For thin or weakly scattering structures, however, part of the light is not scattered. To account for this so-called ballistic light, we need to modify this distribution. Define $E_b$ as the field of the ballistically transmitted light. We can now modify the distribution to have the properties
\begin{align}
    \avg{\xi_i^*(\vk)}&= E_b\delta(\vk-\mathbf{e}_z)\delta_{ij}\\
    \avg{\xi_i^*(\vk_1)\xi_i(\vk_2)}&=\delta(\vk_2-\vk_1)I(\vk_1) +\nonumber\\
    &\quad\abs{E_b}^2\delta(\vk_1-\mathbf{e}_z)\delta(\vk_2-\mathbf{e}_z)\delta_{ij}
\end{align}
with $\mathbf{e}_z$ the unit vector in the $z$-direction, and $j\in x,y,z$ the polarization of the incident beam, and $\delta_{ij}$ is the Kronecker delta. The additional term gives a constant offset in the correlation functions. After normalizing the total average transmission, we find that two of the equations in Table \ref{tab:intensity-functions} need to be modified to
\begin{align}
    1 + (1-\beta)C_x(2\Delta y) \quad&\text{in = $x$, out = $x$}\\
    1 + (1-\beta)C_y(2\Delta y) \quad&\text{in = $y$, out = $y$}
\end{align}
with $\beta = \abs{E_b}^2/(\abs{E_b}^2 + I_0)$ the relative transmission coefficient of the ballistic component. Unsurprisingly, we find that the interference effect due to mirror symmetry vanishes as $\beta$ tends to 0, i.e. when the sample becomes infinitely thin or fully transparent.

\section{Numerical simulations}
A freely available numerical Maxwell solver (WaveSim) is used to simulate optical wave propagation in our designed structures \cite{Osnabrugge2016, WaveSim}. We have designed two structural models; the first model consists of point-like scatterers and the second consists of randomly positioned rods. The main reason to use two different model designs in the simulation is that the point-like scatterers model is appropriate and quick to implement, while the random rods model has more resemblance to our fabricated samples. Since the random rods model is more complex, it requires a more powerful computational system for the simulation. In our models, the structures are mirror symmetric with the symmetry plane the \textit{x}-\textit{z}-plane at \textit{y} = 0. Periodic boundary (PB) conditions are applied in \textit{x}- and \textit{y}-directions, while an anti-reflection boundary layer (ARL), in addition, acyclic convolution (ACC) boundary condition is used in the source injection direction \cite{Osnabrugge2021}. The plane-wave source below the structure injects light with a wavelength of 633 nm along the positive \textit{z}-axis. 
Furthermore, we have used a commercially available finite-difference time-domain (FDTD) solver, Ansys Lumerical, to validate our WaveSim simulation results. In the simulation, we imported the structural model design used in WaveSim and kept all the parameters similar to those used in WaveSim. Again, periodic boundary conditions are applied in \textit{x}- and \textit{y}-directions, while a perfectly matched layer boundary condition is used in the source injection direction. The polarized plane wave with a wavelength of 633 nm is placed below the structure and injects the wave along the positive \textit{z}-axis. The scattered light intensity data is recorded using a 3D frequency-domain field profile monitor.
\section{ Drexhage Analogy }

Drexhage calculated the change in the decay rate of an emitter above a mirror as a consequence of the varying local density of states caused by the interference of the wave emitted by the emitter with the reflection of that wave from the mirror \cite{Drexhage1968,Drexhage1970}. For this calculation, we have calculated the angle-dependent and polarization-dependent emitted field and, for each point on a large far-field sphere, added the fraction of that wave reflected by a virtual zero-phase-shift mirror. Integrating over the sphere resulted in analytic expressions Eqs. 1 and 2 of the main text, yielding the dependence of the intensity on the distance between the emitter and the symmetry plane.
\begin{figure}[!htb]
    \centering
    \makeatletter 
\renewcommand{\thefigure}{S\@arabic\c@figure}
\makeatother
    \includegraphics[width=1\linewidth]{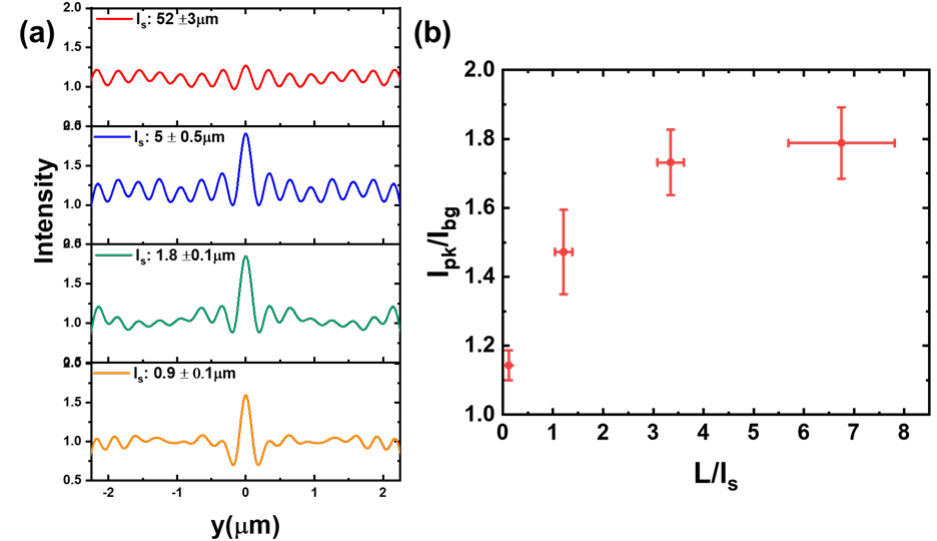}
    \caption{ Numerical simulation results: (a) Intensity line profile variation as a function of $l_s$. The symmetry plane is at \textit{y}=0. Periodic boundary conditions create similar symmetry-induced features at the edges. Intensity enhancement at the symmetry plane versus the diffusion strength (b) the ratio of peak intensity ($I_{pk}$) to the average background intensity ($I_{bg}$) as a function of the ratio of sample thickness and scattering mean free path ($l_s$). The data points in (b) are averaged over 10 configurations.}
    \label{figure}
\end{figure}

The analytic expressions derived in the way of Drexhage assumes an infinitely wide sample and ignores any scattering in the space between the emitter and the mirror or symmetry surface. To investigate the limitations of the simple model we now numerically calculate the intensity for finite-sized samples.  We consider a mirror-symmetric scattering slab having 5\,µm width and 6\,µm thickness. In the simulation, the  positions of the sherical scatterers are fixed, while their refractive index values are varied to get different $l_s$ values. For the ensemble average, ten distinct arrangements of the scatterers are considered; in each arrangement, the scatterers' refractive index is constant, but their placements are altered randomly. We have recorded the transmitted light at the exiting plane of the sample and estimated the ballistic light component. We have evaluated the  $l_s$ values using the Lambert Beer law \(I_b=I_o e^{-L/l_s}\) , where $I_b$ is ballistic light intensity, $I_o$ source intensity, and L is the thickness of the sample. Figure S1(a) shows the intensity line profile variation as a function of $l_s$. It is evident that as the $l_s$ value is reduced, the intensity profile changes. The simulated sample having $l_s=52 \pm 3\,$µm indicates that the light is escaping without interacting with the structure since $l_s \gg L$, and  oscillating fringes around the peak are visible, which we attribute to standing waves because of the periodic boundary conditions. The sample having $l_s=5 \pm 0.5\,$µm suggests weak light scattering caused by structure as $l_s\approx L$. Fringes are still visible around the peak. We hypothesize that they are a combination of the Drexhage-like fringes near the center and some boundary-condition induced Fabry-Perot fringes. For the structures with $l_s \ll L$, the fringes begin to diminish near the peak position. It indicates that only light scattered from scatterers near the symmetry line still constructively interferes at the symmetry plane. From these simulations, we can also determine the ratio of peak intensity ($I_{pk}$) to the average background intensity ($I_{bg}$). Figure S1(b) shows the ratio of $I_{pk} / I_{bg}$ as a function of the ratio of the sample thickness and $l_s$. As the scattering strength of the sample increases, the ratio of $I_{pk} / I_{bg}$ increases, but never reaches the factor 2 predicted by the analytic approximation based on Drexhage's theory.  
\section{Effect of imperfect symmetry}
In the previous sections, we assumed that there was a perfect mirror symmetry in the sample, and that the sample was illuminated with a plane wave perfectly orthogonal to the surface of the sample.
Any deviation from these conditions will reduce the visibility of the symmetry effects.

It is not straightforward to say how manufacturing imperfections affect the symmetry of the sample. The case of single particle deep inside a strongly scattering medium has been studied in detail \cite{cao2022shaping}. However, in finite size samples with a thickness that is not much larger than the transport mean free path, and scatterers with dimensions larger than the wavelength, a general analytical description does not exist.

Therefore, we use a pragmatic approach where we decompose the field coming from the right-hand side of the sample ($E_R$) into a fraction that is a perfect mirror copy of the left-hand side, and a residual term $E_\Delta$, as described in Ref. \cite{vellekoop2008universal}. 
We now arrive at a modified version of Eq.~\eqref{eq:LR-start}
\begin{equation}
    E_R(x,y) = \gamma E_L(x,-y)  + \sqrt{1-\abs{\gamma}^2} E_\Delta(x, y)
\end{equation}
where $\gamma$ ranges from $0$ for a completely disordered medium, to $1$ for perfect symmetry. 
As before, we derive the intensity distribution close to the symmetry plane at the sample surface: 
\begin{align}
    I(x,y) &= \abs{E_L(x,y)}^2 + \abs{\gamma}^2 \abs{E_L(x,-y)}^2  \nonumber\\
    &+ (1-\abs{\gamma}^2)\abs{E_\Delta(x,-y)}^2\nonumber\\
    &+ 2\Re \left[\gamma E_L(x,y)^* E_L(x,-y)\right]
\end{align}
Here, the last term describes the intensity modulation due to the symmetry plane. It can be seen that the visibility of these modulations is proportional to $\gamma$.

Figure S2 presents the estimated $\gamma$ from our numerical simulation results, representing the field correlation between the ideal mirror copy and the perturbed mirror copy of the field. In Figure S2(a), a subset of scatterers was perturbed by distances sampled from a Gaussian distribution with a mean of zero and a standard deviation of 200 nm from their original positions. The resulting speckle patterns for both polarizations were then compared to those from a non-perturbed ensemble. In Figure S2(b), all scatterers in the medium were perturbed with perturbations drawn from Gaussian distributions, with standard deviations incrementally increasing from 40 nm to 200 nm, to evaluate the optical response for both polarizations. Statistical averages are calculated from five realizations for each scenario. 

\begin{figure}[!htb]
    \centering
    \makeatletter 
\renewcommand{\thefigure}{S\@arabic\c@figure}
\makeatother
    \includegraphics[width=1\linewidth]{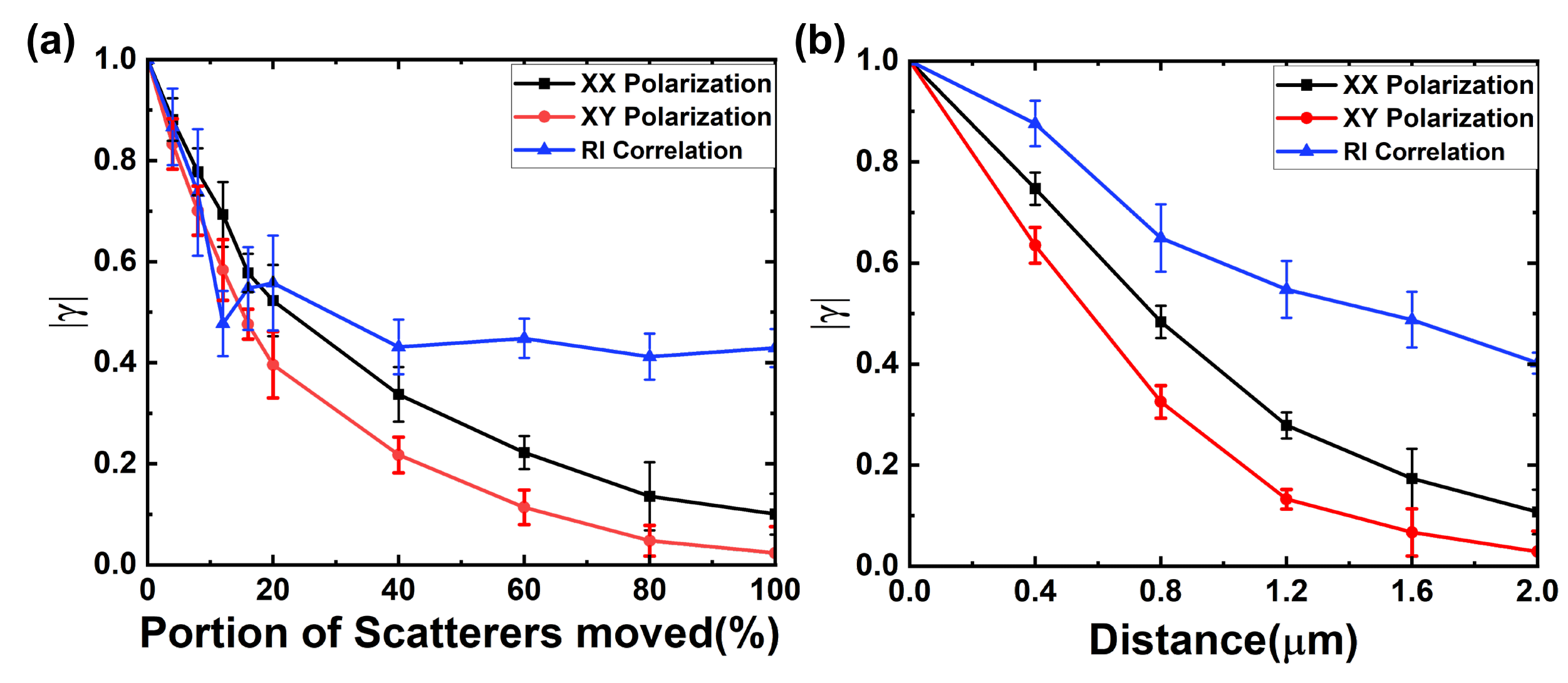}
    \caption{ The $\gamma$ of the electric field for XX and YY polarizations, along with the refractive index distribution correlation, is plotted against the portion of scatterers perturbed by distances drawn from a Gaussian distribution with a standard deviation of 200 nm in the mirror-symmetric medium. Error bars denote the standard deviation, while the solid lines represent the ensemble average values. (b) The $\gamma$ of the electric field and refractive index correlations for scenarios where varying standard deviations perturb all scatterers. The results reveal that the $\gamma$ falls below 50 percent once the average displacement exceeds approximately one-tenth of the wavelength (60 nm for a 600 nm wavelength).}
    \label{figure}
\end{figure}

\section{{Sample fabrication}} 

Following the procedure of \cite{Marakis2020}, a commercial direct laser writing system (Nanoscribe Professional GT) was used to synthesize the structures. A cleaned glass micro slide is taken and a polymer photoresist Nanoscribe IP-G (refractive index of 1.51 for a wavelength of 633\,nm) is drop-cast on it and baked at 120\,°C. The polymer photoresist is a gel, and its high viscosity ensures that the structures' features do not wander during the writing operations, resulting in only minor structural deformation. The coated photoresist is scanned using the focused laser according to the structure coordinates, resulting in solidifying the polymer by 2-photon polymerization. The rods attached to the substrate are written first to create a stable scaffold. This approach minimizes the drifting of features during the writing process. Once the initial scaffold is established, the subsequent rods are written to complete the structure. After the laser writing process, a developer solution is used to wash away the unsolidified polymer photoresist, leaving behind only the desired structure. This process ensures precise control over the position and shape of the scatterers, which is essential for achieving the intended optical properties.
Figure S3 presents the optical imaging outcomes obtained from eight distinct cubes produced by adjusting the writing speed during fabrication. The results demonstrate that the symmetry of the samples remains consistent for both the $I_{XX}$ and $I_{YY}$ polarization scenarios.   

\begin{figure}[!htb]
    \centering\makeatletter 
\renewcommand{\thefigure}{S\@arabic\c@figure}
\makeatother
\includegraphics[width=1\linewidth]{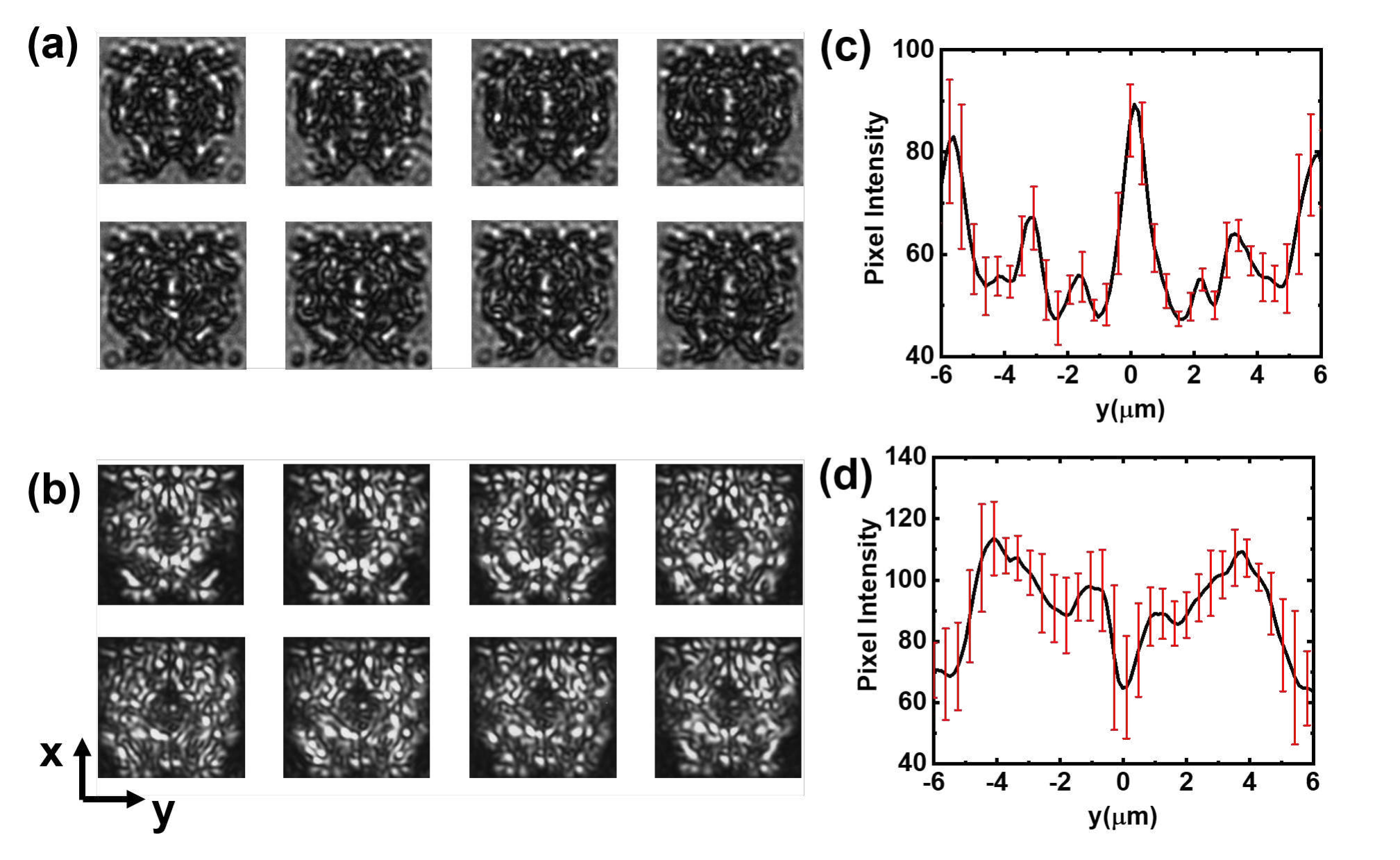}
   \caption{ \textbf{Optical Imaging results:} The measured speckle patterns from eight different cubes in (a) show the $I_{XX}$ component and in (b) the $I_{XY}$ component of the output light. Average intensity line profile summed over the x values are shown in (c) and (d) for the two respective components.}
    \label{fig:enter-label}
\end{figure}

\section{Optical setup}

The samples are characterized in both transmission and reflection mode using the experimental setup as shown in Fig. S4. The setup is built using two different sources: a linearly polarized He–Ne laser ($\lambda = 633\,$nm) and a LED ($\lambda = 628\,$nm). The laser beam is expanded using a combination of two lenses as a beam expander to reduce beam divergence, achieving near-plane-wave illumination of the sample. The sample substrate is placed in a dedicated sample holder and placed into a tilt mount, which is attached to an \textit{x-y-z}-translation stage. Light that is scattered through the sample is collected by an objective with NA = 0.55 and focused onto a CCD camera (Stingray F145 B). The focal length of the objective is: $f_{obj}$ = 4\,mm and the focal lens of the tube lens is: $f_2$ = 200\,mm. This yields a magnification of 50. The camera has a resolution of 1388 x 1038 pixels a pixel size of 6.45\,µm × 6.45\,µm. A magnification of 50x leads to roughly 5 pixels per wavelength, which is adequate for recording clear images. 

\begin{figure}[!htb]
    \centering\makeatletter 
\renewcommand{\thefigure}{S\@arabic\c@figure}
\makeatother
    \includegraphics[width=1\linewidth]{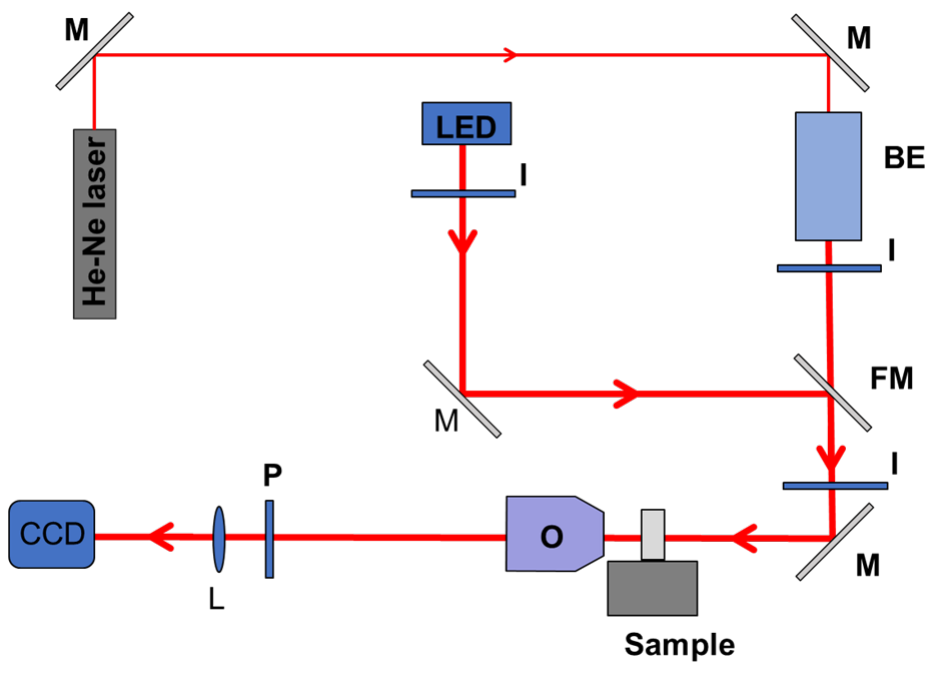}
    \caption{ \textbf{Optical Imaging setup:} Schematic of the optical setup used to characterize the fabricated samples, where M-Mirror, FM-Flip mirror, BE-Beam expander, P-Linear polarizer I- Iris, L-lens, and O-Objective.}
    \label{fig:enter-label}
\end{figure}


\begin{thebibliography}{10}
\bibitem{Gross1996}
D. J. Gross, The role of symmetry in fundamental physics, Proc. Natl. Acad. Sci. U.S.A. \textbf{93}, 14256 (1996).
\bibitem{Joannopoulos2008}
J. D. Joannopoulos, S. G. Johnson, J. N. Winn, and R. D. Meade, Molding the flow of light, (Princeton University Press Princeton, NJ, 2008).
\bibitem{Rotter2017}
S. Rotter and S. Gigan, Light fields in complex media: Mesoscopic scattering meets wave control, Rev. Mod. Phys. \textbf{89}, 015005 (2017).
\bibitem{Ishimaru1978}
A. Ishimaru, Wave propagation and scattering in random media (Academic press New York, 1978), Vol. 2.
\bibitem{Akkermans2007}
E. Akkermans and G. Montambaux, Mesoscopic Physics of Electrons and Photons (Cambridge University Press, Cambridge, 2007).
\bibitem{Wiersma2013}
D. S. Wiersma, Disordered photonics, Nat. Photonics \textbf{7}, 188 (2013).
\bibitem {Jacucci2021}
G. Jacucci, L. Schertel, Y. Zhang, H. Yang, and S. Vignolini, Light Management with Natural Materials: From Whiteness to Transparency, Adv. Mater. \textbf{33}, 2001215 (2021).
\bibitem{ Wiersma1997}
D. S. Wiersma, P. Bartolini, A. Lagendijk, and R. Righini, Localization of light in a disordered medium, Nature \textbf{390}, 671 (1997).
\bibitem{Yu2021}
S. Yu, C.-W. Qiu, Y. Chong, S. Torquato, and N. Park, Engineered disorder in photonics, Nat. Rev. Mater. \textbf{6}, 226 (2021).
\bibitem{Riboli2014}
F. Riboli et al., Engineering of light confinement in strongly scattering disordered media, Nat. Mater. \textbf{13}, 720 (2014).
\bibitem{Liew2011}
S. F. Liew, J.-K. Yang, H. Noh, C. F. Schreck, E. R. Dufresne, C. S. O’Hern, and H. Cao, Photonic band gaps in three-dimensional network structures with short-range order, Phys. Rev. A \textbf{84}, 063818 (2011).
\bibitem{Liew2011}
S. F. Liew et al., Short-range order and near-field effects on optical scattering and structural coloration, Opt. Express \textbf{19}, 8208 (2011).
\bibitem{Saini2020}
S. K. Saini and R. V. Nair, Selective-frequency-gap-induced negative anisotropic scattering in designer photonic structures with short-range order, Phys. Rev. A \textbf{102}, 033529 (2020).
\bibitem{Whitney2009}
R. S. Whitney, P. Marconcini, and M. Macucci, Huge Conductance Peak Caused by Symmetry in Double Quantum Dots, Phys. Rev. Lett. \textbf{102}, 186802 (2009).
\bibitem {Marconcini2013}
P. Marconcini and M. Macucci, Symmetry-dependent transport behavior of graphene double dots, J. Appl. Phys. \textbf{114}, 163708 (2013).
\bibitem{Chéron2019}
É. Chéron, S. Félix, and V. Pagneux, Broadband-enhanced transmission through symmetric diffusive slabs, Phys. Rev. Lett. \textbf{122}, 125501 (2019).
\bibitem{Davy2021}
M. Davy, C. Ferise, É. Chéron, S. Félix, and V. Pagneux, Experimental evidence of enhanced broadband transmission in disordered systems with mirror symmetry, Appl. Phys. Lett. \textbf{119}, 141104 (2021).
\bibitem{Marakis2020}
E. Marakis, R. Uppu, M. L. Meretska, K. J. Gorter, W. L. Vos, and P. W. H. Pinkse, Deterministic and controllable photonic scattering media via direct laser writing, Adv. Opt. Mater. \textbf{8}, 2001438 (2020).
\bibitem{Bender2022}
N. Bender, A. Goetschy, C. W. Hsu, H. Yilmaz, P. J. Palacios, A. Yamilov, and H. Cao, Coherent enhancement of optical remission in diffusive media, Proc. Natl. Acad. Sci. U.S.A. \textbf{119}, e2207089119 (2022).
\bibitem{Osnabrugge2016}
G. Osnabrugge, S. Leedumrongwatthanakun, and I. M. Vellekoop, A convergent Born series for solving the inhomogeneous Helmholtz equation in arbitrarily large media, J. Comput. Phys. \textbf{322}, 113 (2016).
\bibitem{Drexhage1968}
K. Drexhage, H. Kuhn, and F. Schäfer, Variation of the fluorescence decay time of a molecule in front of a mirror, Ber. Bunsenges. Phys. Chem. \textbf{72}, 329 (1968).
\bibitem {Drexhage1970}
K. Drexhage, Influence of a dielectric interface on fluorescence decay time, J. Lumin. \textbf{1}, 693 (1970).
\bibitem {Marakis2019}
E. Marakis, M. C. Velsink, L. J. Corbijn van Willenswaard, R. Uppu, and P. W. H. Pinkse, Uniform line fillings, Phys. Rev. E \textbf{99}, 043309 (2019).
\end{thebibliography}

\begin{thebibliography}{10}
 \bibitem{Osnabrugge2016}
 G. Osnabrugge, S. Leedumrongwatthanakun, and I. M. Vellekoop, A convergent Born series for solving the inhomogeneous Helmholtz equation in arbitrarily large media, J. Comput. Phys. \textbf{322}, 113–124 (2016).

 \bibitem{Osnabrugge2021}
 G. Osnabrugge, M. Benedictus, and I. M. Vellekoop, Ultra-thin boundary layer for high-accuracy simulations of light propagation, Opt. Expr. \textbf{29}, 1649 (2021).

 \bibitem{Drexhage1968}
 K. Drexhage, H. Kuhn, and F. Schäfer, Variation of the fluorescence decay time of a molecule in front of a mirror, Ber. Bunsenges. Phys. Chem. \textbf{72}, 329 (1968).

 \bibitem{Drexhage1970}	
 K. Drexhage, Influence of a dielectric interface on fluorescence decay time, J. Lumin. \textbf{1}, 693 (1970).

 \bibitem{Marakis2020}
E. Marakis, R. Uppu, M. L. Meretska, K. J. Gorter, W. L. Vos, and P. W. H. Pinkse, Deterministic and controllable photonic scattering media via direct laser writing, Adv. Opt. Mater. \textbf{8}, 2001438 (2020).

 \bibitem{WaveSim}
G. Osnabrugge et al., Wavesim - A fast and accurate method for solving the Helmholtz and time-independent Maxwell's equation,  {\url{https://github.com/IvoVellekoop/wavesim}}

 \bibitem{cao2022shaping}
H. Cao, A. P. Mosk, and S. Rotter, Shaping the propagation of light in complex media, Nat. Phys.  \textbf{18}, 994 (2022).

 \bibitem{vellekoop2008universal}
I. M. Vellekoop and A. Mosk, Universal optimal transmission of light through disordered materials, Phys.  Rev. Lett. \textbf{101}, 120601 (2008).

\end{thebibliography}

\end{document}